\documentclass[aps,prd,10pt,a4paper,superscriptaddress,preprintnumbers,showpacs,notitlepage,eqsecnum,nofootinbib]{revtex4-1}

\usepackage[english]{babel}
\usepackage{graphicx}
\usepackage{amsmath}
\usepackage{xcolor}
\usepackage{eufrak}
\usepackage{caption}
\usepackage{subcaption}
\def\square{\mathchoice\sqr54\sqr54\sqr{2.1}3\sqr{1.5}3}
\def\sqr#1#2{{\vcenter{\vbox{\hrule height.#2pt\hbox{\vrule
width.#2pt height#1pt \kern#1pt\vrule width.#2pt}\hrule height.#2pt}}}}
\def\square{\mathchoice\sqr54\sqr54\sqr{2.1}3\sqr{1.5}3}

\makeatletter

\@addtoreset{equation}{section}
\makeatother

\begin{document}

\title{Reducing the two-body problem in scalar-tensor theories\\
to the motion of a test particle :\\
a scalar-tensor effective-one-body approach
}
\author{F\'elix-Louis Juli\'e}

\affiliation{APC, Universit\'e Paris Diderot,\\
CNRS, CEA, Observatoire de Paris, Sorbonne Paris Cit\'e\\
 10, rue Alice Domon et L\'eonie Duquet, F-75205 Paris CEDEX 13, France.}

\date{September 28, 2017}

\begin{abstract}
Starting from the second post-Keplerian (2PK) Hamiltonian describing the conservative part of the two-body dynamics in massless scalar-tensor (ST) theories, we build an effective-one-body (EOB) Hamiltonian which is a $\nu$-deformation (where $\nu= 0$ is the test mass limit) of the analytically known ST Hamiltonian of a test particle. This ST-EOB Hamiltonian leads to a simple (yet canonically equivalent) formulation of the conservative 2PK two-body problem, but also defines a resummation of the dynamics which is well-suited to ST regimes that depart strongly from general relativity (GR) and which
 may provide information on the strong-field dynamics, in particular, the ST innermost stable circular orbit (ISCO) location and associated orbital frequency. Results will be compared and contrasted with those deduced from the ST-deformation of the (5PN) GR-EOB Hamiltonian previoulsy obtained in [Phys. Rev. D\textbf{95}, 124054 (2017)].
\end{abstract}

\maketitle

\section{Introduction\label{introduction}}

Building libraries of accurate gravitational waveform templates is essential for detecting the coalescence of compact binary systems.
To this aim, the effective-one-body (EOB) approach has proven to be a very powerful framework to analytically encompass and combine the post-Newtonian (PN) and numerical descriptions of the inspiral, merger, as well as ``ring-down" phases of the dynamics of binary systems of comparable masses in general relativity, see, e.g., \cite{Damour:2016bks}.\\

Matching and comparing gravitational wave templates to the present and future data from the LIGO-Virgo and forthcoming interferometers will bring the opportunity to test GR at high PN order \textit{and} in the strong field regime of a merger.
A next step to test gravity in this regime is to match gravitational wave data with templates predicted in the framework of modified gravities.
In this context, scalar-tensor (ST) theories with a single massless scalar field have been the most thoroughly studied. For instance, the corresponding dynamics of binary systems is known at 2.5PN order \cite{Mirshekari:2013vb}.\footnote{or, adopting the terminology of \cite{Damour:1992we}, 2.5 post-Keplerian (PK) order, to highlight the fact that (strong) self-gravity effects are encompassed.} What was hence done in \cite{Lang:2013fna}-\cite{Sennett:2016klh} is the computation of ST waveforms at 2PK relative order (although part of this computation requires information on the ST 3PK dynamics, which is, for now, unknown).\\

In that context, the aim of \cite{Julie:2017pkb} (henceforth paper 1) was to go beyond the (yet poorly known) PK dynamics of modified gravities by extending the EOB approach to scalar-tensor theories.
 More precisely, we started from the ST two-body 2PK Lagrangian obtained by Mirshekari and Will \cite{Mirshekari:2013vb} (no spins, nor finite-size, ``tidal" effects) and deduced from it the corresponding centre-of-mass frame 2PK Hamiltonian.
 That two-body 2PK Hamiltonian was then mapped to that of geodesic motion in an effective, ``ST-deformed" metric, which has the important property to reduce to the 1998 Buonanno-Damour EOB metric \cite{Buonanno:1998gg} in the general relativity limit. 
 When extended to encompass the currently best available (5PN) GR-EOB results, the corresponding ST-EOB Hamiltonian of paper 1 is therefore well-suited to test scalar-tensor theories when considered as parametrised corrections to GR.
However, the scope of this GR-centered EOB Hamiltonian is, by construction, restricted to a regime where the scalar field effects are \textit{perturbative} with respect to general relativity.\\

In their 1998 paper, Buonanno and Damour successfully reduced the general relativistic two-body problem to an effective geodesic motion in a static, spherically symmetric (SSS) metric.
In their approach, they ensured that the effective-one-body dynamics is centered on a particular \textit{one-body problem} in general relativity, namely, the geodesic motion of the reduced mass of the system $\mu=m_A m_B/M$ in the Schwarzschild metric produced by a central body, $M=m_A+m_B$, to which it indeed reduces to in the test-mass limit (i.e., $\nu=0$ with $\nu=\mu/M$).
Consequently, the associated predictions were smoothly connected to those of the motion of a test mass in the Schwarzschild metric (which is known exactly),
 ensuring an accurate resummation of the two-body dynamics that could be pushed up to the strong field regime of the last few orbits before plunge.
 
 \vfill\eject
 
 With the same motivation this paper proposes a mapping where the ST-EOB Hamiltonian reduces, in contrast with what was done in paper 1, to the \textit{scalar-tensor} one-body Hamiltonian in the test mass limit, which describes the motion of a test particle in the metric \textit{and scalar field} generated by a central SSS body. Although the conservative dynamics derived from this Hamiltonian and that proposed in paper 1 (and from the Mirshekari-Will Lagrangian) are the same at 2PK order, when taken as being exact, they define different resummations and hence, a priori different dynamics in the strong field regime which is reached near the last stable orbit. In particular, we shall highlight the fact that our new, \textit{ST-centered}, EOB Hamiltonian is well-suited to investigate ST regimes that depart strongly from general relativity.\\
 
The paper is organised as follows :
 In section \ref{sectionOneBody} we present the Hamiltonian describing the motion of a test particle orbiting in the metric and scalar field generated by a central body (when written in Just coordinates) in scalar-tensor theories, henceforth refered as the real one-body Hamiltonian. In order for the paper to be self-contained, in section \ref{sectionJust} we recall the expression of the two-body Hamiltonian in the centre-of-mass frame obtained in paper 1 at 2PK order. In section \ref{section_st_EOB_Hamiltonian}
 we then reduce the two-body problem to an EOB “$\nu$-deformed” version of the ST one-body problem, by means of a canonical transformation and imposing the EOB mapping relation between their Hamiltonians. 
 We finally study the resummed dynamics it defines~; in particular, we compute the innermost stable circular orbit (ISCO) location and associated orbital frequency in the case of Jordan-Fierz-Brans-Dicke theory. Corrections to general relativity ISCO predictions are compared to the results obtained in paper 1.

\section{The scalar-tensor real one-body problem\label{sectionOneBody}}
\subsection{The metric and scalar field outside a SSS body}
In this paper we limit ourselves to the single, massless scalar field case. Adopting the conventions of Damour and Esposito-Far\`ese (DEF, see e.g. \cite{Damour:1992we} or \cite{Damour:1995kt}), the Einstein-frame action reads in vacuum, that is, outside the sources (setting $G_{*}=c= 1$)~:
\begin{equation}
S_{\rm EF}^{\ vac}[g_{\mu\nu},\varphi]=\frac{1}{16\pi }\int d^4x\sqrt{-g}\,\bigg(R-2g^{\mu\nu}\partial_\mu\varphi\partial_\nu\varphi\bigg)\ ,
\label{actionEF}
\end{equation}
where $R$ is the Ricci scalar and $g= \det g_{\mu\nu}$. The vacuum field equations follow :
\begin{subequations}
\begin{align}
&R_{\mu\nu}=2\partial_\mu\varphi\partial_\nu\varphi\ ,\\
&\square\,\varphi=0\ ,
\end{align}
\label{EOFstVac}%
\end{subequations}
where $R_{\mu\nu}$ is the Ricci tensor and $\square\,\varphi=\partial_\mu\left(\sqrt{-g}g^{\mu\nu}\partial_\nu\varphi\right)$.\\

The vacuum, static and spherically symmetric (SSS) solutions to the Einstein-frame field equations (\ref{EOFstVac}), henceforth, real one-body metric $g^*_{\mu\nu}$ and scalar field $\varphi_*$, have a simple analytical expression in Just coordinates (see, e.g., \cite{Coquereaux:1990qs}) :\footnote{In the following, a star ($*$) shall stand for quantities that refer to the real one-body problem.\\}
\begin{subequations}
\begin{align}
ds_*^2&=-D_*dt^2+\frac{d\rho^2}{D_*}+C_*\rho^2(d\theta^2+\sin^2\theta\,d\phi^2)\ ,\\
\text{with}\quad &D_*(\rho)=\left(1-\frac{\mathfrak a_*}{\rho}\right)^{\frac{\mathfrak b_*}{\mathfrak a_*}},\quad C_*(\rho)=\left(1-\frac{\mathfrak a_*}{\rho}\right)^{1-\frac{\mathfrak b_*}{\mathfrak a_*}},
\end{align}
\label{CD}
\end{subequations}
and
\begin{equation}
\varphi_*(\rho)=\varphi_0+\frac{\mathfrak q_*}{\mathfrak a_*}\ln \left(1-\frac{\mathfrak a_*}{\rho}\right)\ ,\label{varphi}
\end{equation}
where $\varphi_0$ is a constant scalar background that must not be considered as an arbitrary integration constant, but rather as imposed, say, by the cosmological environment \cite{Damour:1992kf}\cite{Damour:1993id}, while the other integration constants $\mathfrak a_*$, $\mathfrak b_*$ and $\mathfrak q_*$ have the dimension of a mass and satisfy the constraint :
\begin{equation}
\mathfrak a_*^2=\mathfrak b_*^2+4\mathfrak q_*^2\ .\label{constraintABQ}
\end{equation}

We note that when $\mathfrak{q_*}=0$, i.e. $\mathfrak{a_*}=\mathfrak{b_*}$, the scalar field is a constant, the metric (\ref{CD}) reduces to Schwarzschild's, and Droste and Just coordinates coincide. Note also that pure vacuum (black hole) solutions exhibit singular scalar field and curvature invariants at $\rho=\mathfrak a_*$. For that reason, SSS black holes cannot carry massless scalar ``hair" (thus $\mathfrak{q_*}=0$) and hence do not differ from Schwarzschild's, see e.g. \cite{Hawking:1972qk} and \cite{Damour:1992we}.\\

One easily checks that expanding (\ref{CD})-(\ref{varphi}) at infinity and in isotropic coordinates ($\rho=\bar\rho+\frac{\mathfrak a^*}{2}+\cdots$), the metric and scalar field behave as
\begin{subequations}
\begin{align}
&\bar g^*_{\mu\nu}=\eta_{\mu\nu}+\delta_{\mu\nu}\left(\frac{\mathfrak b_*}{\bar \rho}\right)+\mathcal{O}\left(\frac{1}{\bar\rho^2}\right)\ ,\\
&\varphi_*=\varphi_0-\left(\frac{\mathfrak a_*}{\bar\rho}\right)+\mathcal{O}\left(\frac{1}{\bar\rho^2}\right)\ ,
\end{align}
\label{asympt_reel}%
\end{subequations}
where $\delta_{\mu\nu}$ is the Kronecker symbol.\\

In order to relate the constants of the vacuum solution to the structure of the body generating the fields, we need the Einstein-frame action inside the source,
\begin{equation}
S_{\rm EF}[g_{\mu\nu},\varphi,\Psi]=\frac{1}{16\pi}\int d^4x\sqrt{-g}\,\bigg(R-2g^{\mu\nu}\partial_\mu\varphi\partial_\nu\varphi\bigg)+S_{m}\left[\Psi,\mathcal{A}^2 (\varphi) g_{\mu\nu} \right]\ ,
\label{actionEF}
\end{equation}
where $\mathcal A(\varphi)$ characterizes the ST theory and $\Psi$ generically stands for matter fields, that are minimally coupled to the Jordan metric, $\tilde g_{\mu\nu}\equiv \mathcal{A}^2 (\varphi) g_{\mu\nu}$. The field equations read
\begin{subequations}
\begin{align}
&R_{\mu\nu}=2\partial_\mu\varphi\partial_\nu\varphi+8\pi\left(T_{\mu\nu}-\frac{1}{2}g_{\mu\nu}T\right)\ ,\label{EOFstA}\\
&\square\,\varphi=-4\pi\alpha(\varphi)T\ ,\label{EOFstB}
\end{align}
\label{EOFst}
\end{subequations}
where $T_{\mu\nu}\equiv -\frac{2}{\sqrt{-g}}\frac{\delta S_m}{\delta g^{\mu\nu}}$ is the Einstein-frame energy-momentum tensor of the source, $T\equiv T^\mu_{\phantom{\mu}\mu}$ and where
\begin{equation}
\alpha(\varphi)\equiv \frac{d\ln \mathcal A(\varphi)}{d\varphi}
\end{equation}
measures the universal coupling strength between the scalar field and matter.\\

The constants $\mathfrak b_*$ and $\mathfrak q_*$ can then be matched to the internal structure of the central body through integration of (\ref{EOFstA}) and (\ref{EOFstB}) as
\begin{align}
&\mathfrak b_*=2\int_0^{\rho_0} d^3x\sqrt{-g}\,(-T^{0}{}_{0}+T^{i}{}_{i})\ ,\qquad
\mathfrak q_*=-\int_0^{\rho_0} d^3x\sqrt{-g}\,\alpha(\varphi)\,T\ ,
\label{mass_charge}
\end{align}
where $\rho_0$ denotes the radius of the central body.\footnote{For example, one rewrites (\ref{EOFstB}) as $(\sqrt{-g}g^{\rho\rho}\varphi')'=-4\pi\sqrt{-g}\alpha(\varphi)T$ (where $a'\equiv da/d\rho$) and integrates both sides between the center of the star, where the fields are supposed to be regular, and its radius $\rho_0$. The left-hand-side hence reads $\int_0^{\rho_0}(\sqrt{-g}g^{\rho\rho}\varphi')'d\rho=\sqrt{-g}g^{\rho\rho}\varphi'\vert_{\rho=\rho_0}=\mathfrak q_* \sin\theta$, using the vacuum expressions (\ref{CD}-\ref{varphi}), by continuity  at $\rho=\rho_0$. Hence, one has $\mathfrak q_*=-(4\pi/\sin\theta)\int_0^{\rho_0}d\rho\sqrt{-g}\alpha T=-\int_0^{\rho_0}d\rho\, d\theta\, d\phi \sqrt{-g}\alpha T$, i.e. (\ref{mass_charge}). One similarly obtains $\mathfrak b_*$ through integration of the $t-t$ component of Einstein's equation (\ref{EOFstA}), see \cite{Coquereaux:1990qs} for the details.} The numerical values of these integrals generically depend on the asymptotic value of the scalar field at infinity $\varphi_0$. Indeed, one can for example model a star as a perfect fluid, together with its equation of state. Given some central density and value for the scalar field $\varphi_c\equiv\varphi(\rho=0)$, one integrates (\ref{EOFst}) and the matter equations of motion from the regular center of the body up to $\rho_0$ where the pressure vanishes. The metric and scalar field are then matched to the exterior solution (\ref{CD})-(\ref{varphi}), fixing uniquely $\mathfrak b_*$, $\mathfrak q_*$ and $\varphi_0$ in terms of the central density and $\varphi_c$. When the equation of state and the baryonic number of the star are held fixed, the exterior fields (i.e. $\mathfrak b_*$ and $\mathfrak q_*$) are completely known as functions of $\varphi_c$ only, or, equivalently, of the scalar field value at infinity, $\varphi_0$, see, e.g. \cite{Damour:1993hw} for an explicit computation.\\

 \subsection{Skeletonizing the source of the gravity field}
 In order to clarify the analysis to come in the forthcoming sections, we now ``skeletonize" the body creating the gravity field~; that is, we phenomenologically replace $S_m$ in (\ref{actionEF}) by a point particle action, as was suggested by Eardley in \cite{Eardley}~:
\begin{equation}
S_m^{skel}[X^\mu, g_{\mu\nu}, \varphi]=-\int M_*(\varphi)\,dS\ ,
\label{skeletonAction}
\end{equation}
where $dS=\sqrt{-g_{\mu\nu}dX^\mu dX^\nu}$ and where $X^\mu(S)$ denotes the location of the skeletonized body. The Einstein-frame mass $M_*(\varphi)$ depends on the value of the scalar field at $X^\mu(S)$ (substracting divergent self contributions), on the specific theory and on the body itself (contrarily to (\ref{actionEF}) where the coupling to the scalar field was universal), hence encompassing the effects of the background scalar field on its equilibrium configuration.\footnote{Note that Eardley-type terms do not depend on the local gradients of $g_{\mu\nu}$ and $\varphi$ and hence cannot account for finite-size, ``tidal" effects~; see e.g. \cite{Damour:1998jk}. In this paper, all tidal effects will be neglected.\\} For a discussion on the validity of the skeletonization procedure, see \cite{lesHouches} and \cite{Damour:1992we}.\\

The question adressed now is to relate the function $M_*(\varphi)$ to the parameters describing the exterior solutions, that is $\mathfrak b_*$ and $\mathfrak q_*$, given a scalar field value at infinity $\varphi_0$. The field equations are given by 
\begin{subequations}
\begin{align}
&R_{\mu\nu}=2\partial_\mu\varphi\partial_\nu\varphi+8\pi\left(T_{\mu\nu}-\frac{1}{2}g_{\mu\nu}T\right)\ ,\quad \text{with}\quad T^{\mu\nu}=\int dS\, M_*(\varphi)\frac{\delta^{(4)}(x-X)}{\sqrt{-g}}\frac{dX^\mu}{dS}\frac{dX^\nu}{dS}\ ,\label{tmunuSkel}\\
\text{and}\quad &\square\,\varphi=4\pi\int dS\, M_*(\varphi)A_*(\varphi)\frac{\delta^{(4)}(x-X)}{\sqrt{-g}}\ ,
\end{align}\label{eofSkel}
\end{subequations}
where we introduced the body-dependent function (``capital alpha")
\begin{equation}
A_*(\varphi)\equiv \frac{d\ln M_*(\varphi)}{d \varphi}\ ,
\label{sensi_central}
\end{equation}
which measures the coupling between the skeletonized body and the scalar field.
 Note that because of the body-dependent function $M_*( \varphi)$, the effective scalar field equation is different from (\ref{EOFstB}) with $T^{\mu\nu}$ given in (\ref{tmunuSkel}), because (\ref{EOFstB}) was derived from the universally coupled action (\ref{actionEF}). Note also that since black holes cannot carry scalar hair, $A_*$ must vanish in that case, i.e. $M_*$ must then reduce to a constant and one recovers general relativity.\\

We now solve these equations in the rest-frame of the skeletonized body, setting $\vec X=\vec 0$. Outside it, the metric and scalar field are of the form (\ref{CD}) and (\ref{varphi}). Moreover, solving the field equations (\ref{eofSkel}) perturbatively around the metric and scalar field backgrounds, i.e. $\bar g^*_{\mu\nu}=\eta_{\mu\nu}+h_{\mu\nu}$, $\varphi_*=\varphi_0+\delta \varphi$, in harmonic coordinates $\partial_\mu(\sqrt{-\bar g}\bar g^{\mu\nu})=0$, easily yields, at linear order
\begin{subequations}
\begin{align}
 &\bar g^*_{\mu\nu}=\eta_{\mu\nu}+\delta_{\mu\nu}\left(\frac{2M_*(\varphi_0)}{\bar\rho}\right)+\mathcal O\left(\frac{1}{\bar \rho^2}\right)\ ,\\
 &\varphi_*=\varphi_0-\frac{M_*(\varphi_0)A_*(\varphi_0)}{\bar \rho}+\mathcal O\left(\frac{1}{\bar \rho^2}\right) ,
 \end{align}
 \label{asympt_effect}%
 \end{subequations}
where the $\varphi_0$-dependence of the fields recalls the fact that the skeletonized body is ``sensitive" to the background value of the scalar field in which it is immersed, that is, $\varphi_0$, as already discussed below (\ref{mass_charge}).\footnote{while, as in GR, the asymptotic (constant) metric at infinity can always be ``gauged away" to Minkowski by means of an appropriate coordinate change.\\}\\

Moreover, by comparing (\ref{asympt_effect}) to (\ref{asympt_reel}), one obtains the following relations (knowing that the harmonic and isotropic coordinates identify at linear order)~:
\begin{align}
\mathfrak b_*=2M_*^0\ ,\quad
\mathfrak q_*=M_*^0A_*^0\ ,\quad \mathfrak a_*=2M_*^0\sqrt{1+{\left(A^0_*\right)}^2}\ ,
\label{link_reel_skel}
\end{align}
see (\ref{constraintABQ}), where and from now on, a zero index denotes a quantity evaluated for $\varphi=\varphi_0$.
Hence, by means of the matching conditions (\ref{link_reel_skel}), we have traded the integration constants of the vacuum solution $\mathfrak b_*$ and $\mathfrak q_*$, which are related to the source stress-energy tensor by (\ref{mass_charge}), for their ``skeleton" counterparts, $M_*^0$ and $A_*^0$, which are the values of the function $M_*(\varphi)$ and its logarithmic derivative evaluated at the background $\varphi_0$.

\subsection{The real one-body problem~: the motion of a test particle\\ in the fields of a skeletonized body in ST theories\label{subsection_One_body}}
We now turn to the motion of a self-gravitating test particle $m_*(\varphi)$, coupled to the fields obtained above, i.e., generated by the central body only. The dynamics is described again by an Eardley-type action,
\begin{equation}
S_*[x^\mu]=-\int\,m_*(\varphi_*)\,ds_*\ ,\label{LagrTestTSjustFondam}
\end{equation}
where $ds_*=\sqrt{- g^*_{\mu\nu}dx^\mu dx^\nu}$ and
where $\varphi_*$ and $g^*_{\mu\nu}$ are the real one-body metric and scalar field, given explicitly in Just coordinates in (\ref{CD}), (\ref{varphi}) together with (\ref{link_reel_skel}). Note that the function $m_*(\varphi_*)$ characterizing the particle can be related too to the properties of an extended test body following the steps presented above, but where the scalar environment is not $\varphi_0$ anymore, and is replaced by the value of the scalar field generated by the central body $\varphi_*$, at the location of the test particle, $\varphi_*(x^\mu(s_*))$.\\

To simplify notations it is convenient to replace $m_*(\varphi_*)$ by the rescaled function 
\begin{equation}
V_*(\varphi_*)\equiv \left(\frac{m_*(\varphi_*)}{m_*^0}\right)^2\ ,\quad\text{such that}\quad S_*[x^\mu]=-m_*^0\int\,\sqrt{V_*}\,ds_*\ ,\label{def_V}
\end{equation} 
where we recall that $m_*^0=m_*(\varphi_0)$ is the value of $m_*(\varphi_*)$ when the test particle is infinitely far away from the central body.
Therefore, the scalar-tensor Lagrangian for our test particle, defined as $S_*\equiv\int dt\, L_*$, reads (restricting the motion to the equatorial plane, $\theta=\pi/2$)~:
\begin{equation}
L_*=-m_*^0\sqrt{-(V_*g^*_{\mu\nu})\frac{dx^\mu}{dt}\frac{dx^\nu}{dt}}=-m_*^0\sqrt{V_*\left(D_*-\frac{\dot{\rho}^2}{D_*}-C_*\,\rho^2\dot{\phi}^2\right)}\ ,\quad \quad \dot{\rho}\equiv\frac{d\rho}{dt} \ ,\quad \dot{\phi}\equiv\frac{d\phi}{dt}\ ,
\label{LagrTestTSjust}
\end{equation}
 with
\begin{equation}
D_*(\rho)=\left(1-\frac{a_*}{\hat\rho}\right)^{\frac{b_*}{a_*}},\quad C_*(\rho)=\left(1-\frac{a_*}{\hat\rho}\right)^{1-\frac{b_*}{a_*}},\label{oneBodyDC}
\end{equation}
where we have introduced the dimensionless radial coordinate
\begin{equation}
\hat{\rho}\equiv \rho/M^0_*\ ,
\end{equation}
and where the rescaled constants $b_*$ and $a_*$ follow from (\ref{link_reel_skel}),
\begin{equation}
b_*=2\quad ,\quad a_*=2\sqrt{1+{\left(A^0_*\right)}^2}\ .\label{param_a_b_oneBody}
\end{equation}

In contrast, the expression of $V_*(\varphi_*(\rho))$ (or, equivalently, $m_*$) as an explicit function of $\rho$ depends on the specific ST theory and on the internal structure of the test particle. At 2PK order to which we restrict ourselves in this paper, it will prove sufficient to replace it by its Taylor expansion around $\varphi_0$. To do so, let us introduce the three quantities
\begin{equation}
\alpha_*(\varphi)\equiv\frac{d\ln m_*}{d\varphi}\ ,\qquad
\beta_*(\varphi)\equiv\frac{d\alpha_*}{d\varphi}\ ,\qquad
\beta'_*(\varphi)\equiv\frac{d\beta_*}{d\varphi}\ ,
\label{sensiv_test}
\end{equation}
such that, expanding $m_*(\varphi)$ around $\varphi_0$ (where we recall that $\varphi_0$ is the value at infinity of the scalar field imposed by cosmology) yields
\begin{align}
m_*(\varphi)=m^0_*\left[1+\alpha^0_*(\varphi-\varphi_0)+\frac{1}{2}\bigg({\alpha^0_*}^2+\beta^0_*\bigg)(\varphi-\varphi_0)^2+\frac{1}{6}\bigg(3\beta^0_*\alpha^0_*+{\alpha^0_*}^3+{\beta'}^0_*\bigg)(\varphi-\varphi_0)^3+\cdots\right]\ .\label{massExpansion}
\end{align}
Now, the scalar field generated by the central body is given in (\ref{varphi}) together with (\ref{link_reel_skel}). Hence $V_*$ reads, at 2PK order,
\begin{equation}
V_*(\hat\rho)=\left(\frac{m_*(\varphi_*(\hat\rho))}{m_*^0}\right)^2=1+\frac{v^*_1}{\hat{\rho}}+\frac{v^*_2}{\hat{\rho}^2}+\frac{v^*_3}{\hat{\rho}^3}+\mathcal O\left(\frac{1}{\hat\rho^4}\right) \ ,\label{oneBodyV}
\end{equation}
where the dimensionless constants $v^*_1$, $v^*_2$ and $v^*_3$ depend on the functions $M_*(\varphi)$ and $m_*(\varphi)$ characterizing the central body and the test particle and are given by
\begin{subequations}
\begin{align}
&v^*_1=-2\alpha^0_*A^0_*\ ,\\
&v^*_2=\left(2(\alpha^0_*)^2+\beta^0_*\right)(A^0_*)^2-2\alpha^0_*A^0_*\sqrt{1+(A^0_*)^2}\ ,\\
&v^*_3=-\left(\frac{4}{3}(\alpha^0_*)^3+\frac{1}{3}{\beta'}^0_*+2\alpha^0_*\beta^0_*\right)(A^0_*)^3+\left(4(\alpha^0_*)^2+2\beta^0_*\right)(A^0_*)^2\sqrt{1+(A^0_*)^2}
-\frac{8}{3}\alpha^0_*A^0_*\left(1+(A^0_*)^2\right)\ .
\end{align}
\label{param1corpsTSexact2}
\end{subequations}

To summarize, we have obtained in this section the Lagrangian that describes the dynamics of a test particle orbiting around a central (skeletonized) body in scalar-tensor theories of gravity. At 2PK order, it is entirely described by five coefficients, $a_*$, $b_*$, $v_1^*$, $v_2^*$, $v_3^*$, which are in turn expressed in terms of the five fundamental parameters~: $M_*^0$, $A_*^0$ describing the central body, and $\alpha_*^0$, $\beta_*^0$, ${\beta'}_*^0$ describing the orbiting particle.\footnote{Note that $b_*=\mathfrak b_*/M_*^0$ (with $\mathfrak b_*=2M_*^0$) is a parameter since $M_*^0$ has been factorized out in the definition of $\hat \rho=\rho/M_*^0$.\\}

\section{The real two-body dynamics at 2PK order, a reminder\label{sectionJust}}
In this section, we recall the results from paper 1 \cite{Julie:2017pkb} that will be needed in the forthcoming sections.
\subsection{The two-body 2PK Hamiltonians in scalar-tensor theories\label{subsectionTwo_body_Ham}}
The two-body dynamics is conveniently described in the Einstein-frame (following DEF), by means of an Eardley-type action
\begin{equation}
S_{\rm EF}[x_A^\mu, g_{\mu\nu}, \varphi]=\frac{1}{16\pi}\int d^4x\sqrt{-g}\,\bigg(R-2g^{\mu\nu}\partial_\mu\varphi\partial_\nu\varphi\bigg)-\sum_A\int ds_A\,m_A(\varphi)\ ,
\end{equation}
where $ds_A=\sqrt{- g_{\mu\nu}dx_A^\mu dx_A^\nu}$, and where $x^\mu_A(s_A)$ denotes the position of body $A$. The masses $m_A(\varphi)$ depend on the (regularized) local value of the scalar field and are related to their Jordan-frame counterparts through $m_A(\varphi)\equiv\mathcal A(\varphi) \tilde m_A(\varphi)$.  In the negligible self-gravity limit, the ``Jordan masses" reduce to constants, $\tilde m_A(\varphi)=cst$, so that the motion is a geodesic of the Jordan metric $\tilde g_{\mu\nu}=\mathcal A^2 g_{\mu\nu}$. In contrast, general relativity is recovered when the ``Einstein masses" are constants, $m_A(\varphi)=cst$.\\

We now define a set of body-dependent quantities, consistently with (\ref{sensi_central}) and (\ref{sensiv_test}),
\begin{subequations}
\begin{align}
&\alpha_A(\varphi)\equiv\frac{d\ln m_A}{d\varphi}\quad \left(=\frac{d\ln \mathcal A}{d\varphi}+\frac{d\ln\tilde{m}_A}{d\varphi}\right)\ ,\label{defSensitivitiesA}\\
&\beta_A(\varphi)\equiv\frac{d\alpha_A}{d\varphi}\ ,\label{defSensitivitiesB}\\
&\beta'_A(\varphi)\equiv\frac{d\beta_A}{d\varphi}\ ,\label{defSensitivitiesC}
\end{align}
\label{defSensitivities}%
\end{subequations}
that appear in the 2PK two-body Lagrangian. In the negligible self-gravity limit, $\tilde m_A=cst$, and hence
\begin{equation}
\alpha_A\rightarrow\alpha\equiv\frac{d\ln \mathcal A}{d\varphi}\ ,\quad\beta_A\rightarrow\beta\equiv\frac{d\alpha}{d\varphi}\ ,\quad\beta'_A\rightarrow\beta'\equiv\frac{d\beta}{d\varphi}\ ,\label{limit_noSelfGrav}
\end{equation}
become universal, while in the general relativity limit, $m_A=cst$, implying $
\alpha_A=\beta_A=\beta'_A=0$.\\

The conservative part of the scalar-tensor two-body problem has been studied at 1PK order by Damour and Esposito-Far\`ese (DEF) in \cite{Damour:1992we} and at 2PK order by DEF in \cite{Damour:1995kt} and Mirshekari and Will (MW) in \cite{Mirshekari:2013vb}, performing a small orbital velocities, weak field expansion ($V^2\sim m/R$) around $\eta_{\mu\nu}$ and a constant cosmological background $\varphi_0$. Because of the harmonic coordinates in which it has been computed, the two-body Lagrangian depends linearly on the accelerations of the bodies at 2PK level.

\vfill\eject

In paper 1, we started from this MW Lagrangian, $L\left(\vec Z_{A/B},\dot{\vec Z}_{A/B},\ddot {\vec Z}_{A/B}\right)$. Once translated in terms of the DEF conventions presented above (see also paper 1, appendix A), we eliminated the dependence in the accelerations $\ddot {\vec Z}_{A/B}$ by means of suitable contact transformations of the form
\begin{equation}
\vec Z'_A(t)=\vec Z_A(t)+\delta\vec Z_A\left(\vec Z_{A/B},\dot{\vec Z}_{A/B}\right)\ ,
\end{equation}
that is, four-dimensional 2PK coordinate changes. We found a whole class of coordinate systems, labeled by fourteen parameters $f_i$, in which the Lagrangian is ordinary (see paper 1 appendix B and below). By means of a further Legendre transformation, we obtained the associated Hamitonians $H(Q,P)$ in the center-of-mass frame, the conjugate variables being $\vec Z=\vec Z_A-\vec Z_B$ and $\vec P=\vec P_A=-\vec P_B$, and in polar coordinates~: $(Q,P)\equiv(R,\Phi, P_R, P_\Phi)$ where $P_R=\vec N\cdot\vec P$ and $P_\Phi=R(\vec N\times\vec P)_z$. The resulting isotropic, translation-invariant, ordinary Hamiltonians are given at 2PK order in paper 1, section III C,
\begin{equation}
\hat{H}\equiv \frac{H}{\mu}=\frac{M}{\mu}+\left(\frac{\hat{P}^2}{2}-\frac{G_{AB}}{\hat{R}}\right)+\hat H^{\rm 1PK}+\hat H^{\rm 2PK}+\cdots
\label{ham2corpsStructure}
\end{equation}
where we have introduced the rescaled quantities
\begin{equation}
\hat P^2\equiv \hat P_R^2+{\hat P_\Phi^2\over \hat R^2}\quad\hbox{with}\quad \hat P_R\equiv\frac{P_R}{\mu}\ ,\ \hat P_\Phi\equiv{P_\Phi\over\mu M}\ ,\ \hat R\equiv {R\over M}\ ,
\label{impulsReduites}
\end{equation}
and the reduced mass, total mass and symmetric mass ratio :
\begin{equation}
\mu\equiv{m_A^0m_B^0\over M}\ ,\ M\equiv m_A^0+m_B^0\ ,\ \nu\equiv{\mu\over M}\ ,\label{reducedTotalMasses}
\end{equation}
where $m_A^0$ and $m_B^0$ are the values of the functions $m_A(\varphi)$ and $m_B(\varphi)$ at $\varphi=\varphi_0$.\\

At 2PK order, the two-body Hamiltonians depend on seventeen coefficients $(h_i^{\,\rm n PK})$ (which are very lenghty and are given explicitely in appendix C of paper 1), which in turn depend on the fourteen $f_i$ parameters and on the eleven following combinations of the eight fundamental mass parameters (\ref{defSensitivities}) [$m_A^0$, $\alpha_A^0$, $\beta_A^0$ and $\beta_A^{'0}$ and B counterparts, characterizing at 2PK order the functions $m_{A/B}(\varphi)$]~:
\begin{subequations}
\begin{align}
&m_A^0\ ,\quad  G_{AB}\equiv 1+\alpha_A^0\alpha_B^0\ ,\label{defParamPK1}\\
&\bar\gamma_{AB}\equiv -\frac{2\alpha_A^0\alpha_B^0}{1+\alpha_A^0\alpha_B^0}\ ,\quad
\bar\beta_A\equiv\frac{1}{2}\frac{\beta_A^0(\alpha_B^0)^2}{(1+\alpha_A^0\alpha_B^0)^2}\ ,
\label{defParamPK2}\\
&\delta_A\equiv\frac{(\alpha_A^0)^2}{(1+\alpha_A^0\alpha_B^0)^2}\ ,\quad\epsilon_A\equiv\frac{(\beta'_A\alpha_B^3)^0}{(1+\alpha_A^0\alpha_B^0)^3}\ ,\quad\zeta\equiv\frac{\beta_A^0\alpha_A^0\beta_B^0\alpha_B^0}{(1+\alpha_A^0\alpha_B^0)^3}\ ,\label{defParamPK3}
\end{align}
\label{defParamPK}%
\end{subequations}
and $(A\leftrightarrow B)$ counterparts, where we recall that a zero index indicates a quantity evaluated at infinity, $\varphi=\varphi_0$. In the general relativity limit, $m_A=cst$, the Hamiltonian considerably simplifies since these combinations reduce to
\begin{equation}
G_{AB}=1\ ,\quad\text{and}\quad \bar\gamma_{AB}=\bar\beta_A=\delta_A=\epsilon_A=\zeta=0\ .\label{GR_limit}
\end{equation}

\subsection{The canonical transformation\label{section_transfoCano}}
The EOB mapping consists in imposing a functional relation between the two-body Hamiltonian $H(Q,P)$, and an effective Hamiltonian $H_e$ (that we shall build in the next section), by means of a canonical transformation,
\begin{equation}
(Q,P)\rightarrow(q,p)\ ,
\end{equation}
where $(q,p)\equiv(\rho,\phi,p_\rho,p_\phi)$. The canonical transformation is generated by the (time-independent and isotropic) generic function $G(Q,p)$ introduced in \cite{Julie:2017pkb}, section III D, which depends on nine parameters at 2PK order,
\begin{equation}
{G(Q,p)\over \mu M}=\hat R\, \hat p_\rho\left[\bigg(\alpha_1{\cal P}^2+\beta_1\hat p_\rho^2+{\gamma_1\over\hat R}\bigg)+\bigg(\alpha_2{\cal P}^4+\beta_2{\cal P}^2\hat p_\rho^2+\gamma_2\hat p_\rho^4+\delta_2{{\cal P}^2\over \hat R}+\epsilon_2{\hat p_\rho^2\over \hat R}+{\eta_2\over \hat R^2}\bigg)+\cdots\right]\ ,\label{generatrice}
\end{equation}
where we introduced the reduced quantities
\begin{equation}
\quad {\cal P}^2\equiv\hat p_\rho^2+{\hat p_\phi^2\over\hat R^2}\ ,\quad \hat R\equiv {R\over M}\ ,\quad\hat p_\rho\equiv\frac{p_\rho}{\mu}\ ,\quad \hat p_\phi\equiv\frac{p_\phi}{\mu M}\ .
\end{equation}
The associated canonical transformation reads
\begin{equation}
\rho(Q,p)=R+{\partial G\over\partial p_\rho}\ ,\quad \phi(Q,p)=\Phi+{\partial G\over\partial p_\phi}\ ,\quad P_R(Q,p)=p_\rho+{\partial G\over\partial R}\ ,\quad P_\Phi(Q,p)=p_\phi+{\partial G\over\partial\Phi}\ ,
\label{transfoCano}
\end{equation}
and leads to 1PK and higher order coordinate changes. Note that the $\Phi$-independence of $G(Q,p)$ yields $P_\Phi=p_\phi$. Moreover, for circular orbits, $p_\rho=0\Leftrightarrow P_R=0$, we note that $\phi=\Phi$ and hence only the radial coordinates differ $\rho\neq R$.\\

The two-body Hamiltonian (\ref{ham2corpsStructure}) is thus rewritten in the intermediate coordinate system $H'(Q,p)=H(Q,P(Q,p))$ using the last two equations in (\ref{transfoCano}) which yield (dropping the prime)
\begin{equation}
\hat H=\frac{M}{\mu}+\left(\frac{{\cal P}^2}{2}-\frac{G_{AB}}{\hat R}\right)+\hat H^{\rm 1PK}+\hat H^{\rm 2PK}+\cdots\ ,\label{H_cano}
\end{equation}
where the explicit expressions for $\hat H^{\rm 1PK}$ and $\hat H^{\rm 2PK}$ are given in appendix D of paper 1. It depends on the eight fundamental parameters (\ref{defSensitivities}), on the fourteen parameters $f_i$ characterizing the coordinate system in which the two-body Hamiltonian $H(Q,P)$ was written, and on the nine parameters of the canonical transformation (\ref{generatrice}).

\section{The scalar-tensor EOB Hamiltonian\label{section_st_EOB_Hamiltonian}}
In this section we relate the canonically transformed, two-body Hamiltonians $H(Q,p)$ to the Hamiltonian $H_e$ of an effective test-particle in the fields of an effective central body.\\

To this aim, we shall propose a ST-centered Hamiltonian $H_e$ that contrasts with what was done in paper 1, where $H_e$ was centered on the GR limit.

\subsection{The effective Hamiltonian\label{subs_effect_Hamiltonian}}

In view of reducing the two-body dynamics to that of an effective test particle coupled to the generic SSS fields of an effective single body, and taking inspiration from (\ref{LagrTestTSjustFondam}), let us consider the action (setting again $\theta=\pi/2$) :
\begin{equation}
S_e[x^\mu]=-\int m_e(\varphi_e)\,ds_e
\end{equation}
where $ds_e=\sqrt{-g^e_{\mu\nu}dx^\mu dx^\nu}$ and where $x^\mu[s_e]$ is the world-line of the effective particle characterized by the function $m_e(\varphi_e)$. As in (\ref{LagrTestTSjust}), we write the effective metric in Just coordinates
\begin{equation}
ds_e^2=-D_e\,dt^2+\frac{d\rho^2}{D_e}+C_e\,\rho^2d\phi^2\ ,
\end{equation}
where $D_e$ and $C_e$ are effective functions to be determined later.\\

We now replace, for notational convenience, $m_e(\varphi_e)$ by the function
\begin{equation}
V_e\equiv \left(\frac{m_e(\varphi_e)}{\mu}\right)^2\ ,\quad\text{such that}\quad S_e[x^\mu]=-\mu\int \sqrt{V_e}\,ds_e\ ,
\end{equation}
which is the third effective function to be determined, and where $\mu$ is identified to the real two-body reduced mass, defined in (\ref{reducedTotalMasses}).
The associated Lagrangian, defined as $S_e\equiv\int dt\,L_e$, reads therefore 
\begin{equation}
L_e=-\mu\sqrt{-(V_e g^e_{\mu\nu})\frac{dx^\mu}{dt}\frac{dx^\nu}{dt}}=-\mu\sqrt{V_e\left(D_e-\frac{\dot{\rho}^2}{D_e}-C_e\rho^2\dot{\phi}^2\right)}\ ,\quad\text{where}\quad \dot \rho\equiv d\rho/dt\ ,\quad\dot\phi\equiv d\phi/dt\ .
\label{effectiveProblemJust}
\end{equation}
 Note that $L_e$ identifies to the Lagrangian of a geodesic in the \textit{body-dependent} conformal metric, $(V_eg^e_{\mu\nu})$.
 
 \vfill\eject

One easily deduces the effective momenta and Hamiltonian, 
\begin{equation*}
p_\rho\equiv{\partial L_e\over\partial\dot \rho}\ ,\quad p_\phi\equiv{\partial L_e\over\partial\dot\phi}\ ,\quad H_e\equiv p_\rho\dot \rho+p_\phi\dot\phi-L_e\ ,
\end{equation*}
that is
\begin{equation}
\hat H_e\equiv\frac{H_e}{\mu}=\sqrt{V_eD_e+D_e^2\hat p_\rho^2+\frac{D_e}{C_e}\frac{\hat p_\phi^2}{\hat \rho^2}}\ ,
\label{HeffectifReduit}
\end{equation}
where we used the reduced (dimensionless) variables
\begin{equation}
\hat \rho\equiv{\rho\over M}\ ,\quad\hat p_\rho\equiv{p_\rho\over \mu}\ ,\quad\hat p_\phi\equiv{p_\phi\over \mu M}\ ,\quad \hat p^2\equiv\hat p_\rho^2+{\hat p_\phi^2\over \hat \rho^2}\ ,
\label{varChap2}
\end{equation}
$M$ being identified to the real total mass, see (\ref{reducedTotalMasses}).\\

In order to relate the effective Hamiltonian $H_e$ to the two-body (perturbative) Hamiltonian $H$, we now restrict $H_e$ to 2PK order also. To this end, one could in principle expand $V_e$, $D_e$ and $C_e$ in the form of $1/\hat\rho$ series. However, our aim being to build an effective dynamics as close as possible to the \textit{scalar-tensor} test-body problem, we shall rather introduce the non perturbative, ``resummed" ansatz for the metric functions $D_e$ and $C_e$~:
\begin{align}
D_e(\rho)\equiv&\left(1-\frac{a}{\hat\rho}\right)^{\frac{b}{a}},\quad C_e(\rho)\equiv\left(1-\frac{a}{\hat\rho}\right)^{1-\frac{b}{a}},\label{ansatzJustMetric}
\end{align}
as suggested by (\ref{oneBodyDC}), and where $a$ and $b$ are two effective parameters that we shall determine in the following. As already remarked below equation (\ref{effectiveProblemJust}), the effective dynamics is equivalent to the geodesic motion in the conformal metric $(V_eg^e_{\mu\nu})$. The ansatz (\ref{ansatzJustMetric}) that we shall use rather than a simple $1/\hat\rho$ expansion of $D_e$ and $C_e$ is hence crucial, since the latter would be equivalent, to within a mere coordinate change ($r^2=C_eV_e\rho^2$), to the GR-centered approach of paper 1.\\

In contrast, a specific ansatz for the function $V_e$ can be proposed in the framework of a specific ST theory and when the internal structure of the two real bodies is known, see discussion below (\ref{param_a_b_oneBody}). (For an example, see subsection \ref{subsec_JFBD}.)
For the moment, we hence expand $V_e$ a 2PK order, similarly to what was done in (\ref{oneBodyV})~:
\begin{equation}
V_e(\rho)=1+\frac{v_1}{\hat{\rho}}+\frac{v_2}{\hat{\rho}^2}+\frac{v_3}{\hat{\rho}^3}+\cdots\ ,\label{ansatzJustV}
\end{equation}
where $v_1$, $v_2$ and $v_3$ are three further effective parameters to determine later.\\

Expanding the effective Hamiltonian (\ref{HeffectifReduit}) and (\ref{ansatzJustMetric}-\ref{ansatzJustV}) hence reads
\begin{equation}
\hat{H}_e=1+\hat H^{\rm K}_e+\hat H^{\rm 1PK}_e+\hat H^{\rm 2PK}_e+\cdots
\label{HamEffectJust2PK}
\end{equation}
with, at 1PK,
\begin{align}
&\hat H^{\rm K}_e=\frac{\hat p^2}{2}+\frac{v_1-b}{2\hat \rho}\ ,\quad
H^{\rm 1PK}_e=-\frac{\hat p^4}{8}+\frac{1}{4 \hat \rho}\left[\hat p^2 (2 a-3 b-v_1)-2 a \hat p_\rho^2\right]+\frac{1}{8 \hat \rho^2}\left[-2 a b+b^2-2 b v_1-v_1^2+4 v_2\right]\ ,
\end{align}
and, at 2PK,
\begin{align}
&H^{\rm 2PK}_e=\frac{\hat p^6}{16}+\frac{1}{16 \hat \rho}\left[\hat p^4 (5 b+3 v_1-4a)+4 a \hat p^2 \hat p_\rho^2\right]\nonumber\\
&+\frac{1}{16 \hat \rho^2}\left[4 a \hat p_\rho^2 (-2 a + 3 b + v_1) + (8 a^2 + 9 b^2 + 6 b v_1 + 3 v_1^2 - 
    2 a (9 b + 2 v_1) - 4 v_2) \hat p^2\right]\nonumber\\
    &+\frac{1}{48\hat \rho^3}\left[-8 a^2 b - b^3 + 6 a b (b - v_1) + 3 b^2 v_1 + 3 b (v_1^2 - 4 v_2) + 
 3 (v_1^3 - 4 v_1 v_2 + 8 v_3)\right]\ .
\end{align}

In order to relate the two-body Hamiltonians of the previous section \ref{section_transfoCano} and the present effective Hamiltonian $H_e(q,p)$, we finally express the latter in the same coordinate system $H'_e(Q,p)=H_e(q(Q,p),p)$ using the first two relations in (\ref{transfoCano}). The resulting effective Hamiltonian reads (dropping again the prime)
\begin{equation}
\hat H_e=1+\left(\frac{\mathcal P^2}{2}+\frac{v_1-b}{2\hat R}\right)+\hat H_e^{\rm 1PK}+\hat H_e^{2\rm PK}+\cdots\label{He_cano}
\end{equation}
where we recall that ${\cal P}^2\equiv\hat p_\rho^2+\hat p_\phi^2/\hat R^2$ and where $H_e^{\rm 1PK}$ and $H_e^{\rm 2PK}$ are explicitly given in appendix \ref{appendix_canoTransf_He} of this paper.

\subsection{The EOB mapping\label{subsectionMapping}}
By means of the generic canonical transformation (\ref{transfoCano}-\ref{generatrice}), the real and (a priori independent) effective Hamiltonians $H(Q,p)$ and $H_e(Q,p)$ have been written in a common coordinate system, $(Q,p)$ ; see (\ref{H_cano}) and (\ref{He_cano}).
 Now, as discussed in e.g. \cite{Buonanno:1998gg}, \cite{Damour:2000we} and  \cite{Damour:2015isa}, and as proven to be indeed necessary at all orders in GR as well as in ST theories in  \cite{Damour:2016gwp}, both Hamiltonians shall be related by means of the quadratic functional relation (we recall that $\nu=\mu/M$)~:
\begin{equation}
\frac{H_e(Q,p)}{\mu}-1=\left(\frac{H(Q,p)-M}{\mu}\right)
\left[1+\frac{\nu}{2}\left(\frac{H(Q,p)-M}{\mu}\right)\right]\ .
\label{EOBquadrRel}%
\end{equation}
The identification (\ref{EOBquadrRel}) proceeds order by order and term by term to yield a \textit{unique} solution for $H_e$, that is for the funtions introduced in the previous subsection
\begin{align}
D_e(\rho)\equiv&\left(1-\frac{a}{\hat\rho}\right)^{\frac{b}{a}},\quad C_e(\rho)\equiv\left(1-\frac{a}{\hat\rho}\right)^{1-\frac{b}{a}},\quad
V_e(\rho)=1+\frac{v_1}{\hat{\rho}}+\frac{v_2}{\hat{\rho}^2}+\frac{v_3}{\hat{\rho}^3}+\cdots\ ,\label{rappel_D_C_V}
\end{align}
whose effective parameters now depend on the combinations (\ref{defParamPK}) and are the main technical result of this paper~:
\begin{subequations}
\begin{align}
&b=2\ ,\label{coeffsEffectifsJust1}\\
& v_1=-2\alpha_A^0\alpha_B^0\ ,\nonumber\\ \nonumber\\
&a=2 \mathcal R\ ,\\
&v_2= 2 - 4   G_{AB} + 2   \left(1 + \langle\bar\beta\rangle\right)G_{AB}^2 -2\alpha_A^0\alpha_B^0 \mathcal R\ ,\nonumber\\ \nonumber\\
&\frac{v_3}{4}=1 - \frac{5}{3} G_{AB} + \left(1 +  \langle\bar\beta\rangle + 
    \frac{2}{3} \langle\delta\rangle\right) G_{AB}^2 
    - \frac{1}{3} \left(1 + 
    3 \langle\bar\beta\rangle +\frac{1}{4} \langle\epsilon\rangle + 
    2 \langle\delta\rangle\right) G_{AB}^3 
    + 
  \bigg(1 - 2  G_{AB} + \left(1 + \langle\bar\beta\rangle\right) G_{AB}^2  \bigg)\mathcal R\nonumber\\
    &\quad + 
 \nu \left[ \frac{17}{3} G_{AB} - 
    \frac{1}{3} \bigg(19 + 4 \langle\bar\beta\rangle+6\zeta\bigg) G_{AB}^2 + \left(\frac{2}{3} - \frac{3}{4} (\bar\beta_A+\bar\beta_B) + 
       \frac{1}{12} (\epsilon_A +\epsilon_B)\right.\right. 
       \left.\left.+ \frac{1}{6} (\delta_A + \delta_B) + \frac{3}{2} \langle\bar\beta\rangle\right) G_{AB}^3 \right]\ , 
\end{align} \label{coeffsEffectifsJust}
\end{subequations}
where we have introduced
\begin{equation}
\mathcal R\equiv \sqrt{1+\langle\delta\rangle G_{AB}^2+\nu \bigg[8G_{AB}-2  \bigg(1+\langle\bar\beta\rangle\bigg)G_{AB}^2\bigg]}\ ,\label{mathcalR}
\end{equation}
and the ``mean" quantities
\begin{align}
\langle\bar\beta\rangle\equiv \frac{m_A^0\bar\beta_B+m_B^0\bar\beta_A}{M}\ ,\quad
\langle\delta\rangle\equiv\frac{m_A^0 \delta_A + m_B^0 \delta_B}{M}\ ,\quad
\langle\epsilon\rangle\equiv\frac{m_A^0 \epsilon_B + m_B^0 \epsilon_A}{M}\ .\label{meanOverBodiesParam}
\end{align}

We note that as they should, these parameters can alternatively be deduced from the effective metric found in paper 1, using the 2PK-expanded coordinate change $r^2=C_eV_e\rho^2$, where $r$ is the Schwarschild-Droste coordinate used there.\footnote{Note also that the present results (\ref{coeffsEffectifsJust}-\ref{meanOverBodiesParam}) have been simplified using the relation $\bar\gamma_{AB}=-2+2/G_{AB}$, relating $\bar\gamma_{AB}$ to the \textit{dimensionless} combination $G_{AB}$, see (\ref{defParamPK}). The reader willing to establish $G_*$ (i.e. Newton's constant) again should note that it only appears through $\hat\rho\equiv \rho/(G_*M)$.}\\

As a first consistency check, we note that the effective coefficients (\ref{coeffsEffectifsJust}) do not depend on the $f_i$ parameters introduced in section \ref{subsectionTwo_body_Ham}, i.e., on the coordinate system $(R,\Phi)$ in which the two-body Hamiltonian has been initially written, as expected by covariance of the theory. Indeed, the $f_i$ parameters are absorbed in the 2PK part of the canonical transformation (\ref{generatrice}), whose parameters read
\vfill\eject
\begin{align}
&\alpha_1=-\frac{\nu}{2}\ ,\quad\beta_1=0\ ,\quad\gamma_1=G_{AB}\left[\frac{1}{2}\nu+\left(1+\frac{1}{2}\bar\gamma_{AB}\right)\mathcal{R}\right]\ ,\quad\alpha_2=\frac{1}{8}(1-\nu)\nu\ ,\quad\beta_2=0\ ,\quad\gamma_2=\frac{\nu^2}{2}\ ,\nonumber\\
&\delta_2=G_{AB}\left[ f_6 \frac{m_A^0}{M} + f_1 \frac{m_B^0}{M} - 
 \nu \left(f_1 + f_6 +(-f_3+f_5 + f_6) \frac{m_A^0}{M}+ (f_1 + f_2 - f_4) \frac{m_B^0}{M} -\frac{3}{2}-\bar\gamma_{AB}+\frac{\nu}{8} \right)\right]\ ,\nonumber\\
&\epsilon_2=G_{AB}\left[-\frac{\nu^2}{8} + f_{10} \frac{m_A^0}{M} + f_7 \frac{m_B^0}{M} - 
 \nu \bigg(f_7 + f_{10} + (f_9 + f_{10}) \frac{m_A^0}{M} + (f_7 + f_8) \frac{m_B^0}{M}\bigg)\right]\ ,\nonumber\\
&\eta_2= G_{AB}^2\left[f_{13}\frac{m_A^0}{M}+f_{12}\frac{m_B^0}{M}+ \nu \bigg(f_{11}- f_{12}- f_{13}+ f_{14}\bigg) + 
 \nu  \left(-\frac{7}{4}-\bar\gamma_{AB} - \langle\bar\beta\rangle + \frac{\bar\beta_A + \bar\beta_B}{2} + \frac{\nu}{4}\right)\right]\ .\label{coeffCanoJust}
\end{align}

The real two-body Hamiltonian (\ref{ham2corpsStructure}), whose full expression is relegated to section III C and appendix C of paper 1, has hence been reduced to a compact effective Hamiltonian, where most of the two-body Hamiltonian complexity  is hidden in the canonical transformation (\ref{generatrice}), (\ref{coeffCanoJust}) (e.g., information regarding the initial coordinate system) and in the mapping relation (\ref{EOBquadrRel}).

\subsubsection{The $\nu=0$ limit}

Setting formally $\nu=0$ in (\ref{coeffsEffectifsJust}-\ref{mathcalR}), the parameters reduce to, when written in terms of the fundamental quantities (\ref{defSensitivities})~:
\begin{subequations}
\begin{align}
&b=2\ ,\nonumber\\
&v_1=-2\alpha_A^0\alpha_B^0\ ,\\ \nonumber\\
&a=2 \mathcal R\ ,\\ 
&v_2= 2(\alpha_A^0\alpha_B^0)^2+\frac{(m_A\alpha_A^2)^0\beta_B^0+(m_B\alpha_B^2)^0\beta_A^0}{M}-2\alpha_A^0\alpha_B^0\mathcal R\ ,\nonumber\\ \nonumber\\
&v_3=-\frac{4}{3}(\alpha_A^0\alpha_B^0)^3-\frac{1}{3}\frac{(m_A\alpha_A^3)^0{\beta'}_B^0+(m_B\alpha_B^3)^0{\beta'}_A^0}{M}-2\alpha_A^0\alpha_B^0\frac{(m_A\alpha_A^2)^0{\beta}_B^0+(m_B\alpha_B^2)^0{\beta}_A^0}{M}\\
&\quad -\frac{8}{3}\left(1+\frac{(m_A\alpha_A^2)^0+(m_B\alpha_B^2)^0}{M}\right)\alpha_A^0\alpha_B^0+\left(4(\alpha_A^2\alpha_B^2)^0+2\frac{(m_A\alpha_A^2)^0{\beta}_B^0+(m_B\alpha_B^2)^0{\beta'}_A^0}{M}\right)\mathcal{R}\ ,\nonumber\\
&\hspace*{2,5cm}\text{with}\quad\mathcal{R}=\sqrt{1+\frac{(m_A\alpha_A^2)^0+(m_B\alpha_B^2)^0}{M}}\ .\nonumber
\end{align} \label{paramEffectFund} %
\end{subequations}

Identifying now (\ref{paramEffectFund}) to the parameters (\ref{param_a_b_oneBody}) and (\ref{param1corpsTSexact2}) of the real one-body problem presented in section \ref{subsection_One_body} does yield a unique solution~:
\begin{subequations}
\begin{align}
&(A^0_*)^2=\frac{m_A^0(\alpha_A^0)^2+m_B^0(\alpha_B^0)^2}{m^0_A+m_B^0}\ ,\label{AlphaEffectif}\\
&\alpha^0_*=\frac{\alpha_A^0\alpha_B^0}{A^0_*}\ ,\\
&\beta^0_*=\frac{(m_A\alpha_A^2)^0\beta_B^0+(m_B\alpha_B^2)^0\beta_A^0}{(m_A\alpha_A^2)^0+(m_B\alpha_B^2)^0}\ ,\\
&{\beta'}^0_*=\frac{(m_A\alpha_A^3)^0{\beta'}^0_B+(m_B\alpha_B^3)^0{\beta'}_A^0}{(m^0_A+m_B^0)(A^0_*)^3}\ ,
\end{align}
\label{nouveauxParamEffectifs}%
\end{subequations}
together with $m^0_*=\mu$, $M^0_*=M$.\\

We hence conclude that the dynamics described by $H_e$ is a $\nu$-deformation of a scalar-tensor test-body problem, describing an effective test particle characterized by
\begin{equation}
\ln m_*(\varphi)=\ln m_*^0+\alpha_*^0(\varphi-\varphi_0)+\beta_*^0(\varphi-\varphi_0)^2+{\beta'}_*^0(\varphi-\varphi_0)^3+\cdots\ ,\label{effectmu}
\end{equation}
orbiting around an effective central body characterized by
\begin{equation}
\ln M_*(\varphi)=\ln M_*^0+A_*^0(\varphi-\varphi_0)+\cdots\ ,\label{effectM}
\end{equation}
 whose fundamental parameters 
[$M^0_*$, $A^0_*$, $m^0_*$, $\alpha^0_*$, $\beta^0_*$ and ${\beta'}^0_*$] are related to the real, two-body ones through (\ref{nouveauxParamEffectifs}).
Since $\nu\rightarrow 0$ means, say, $m_B^0>>m_A^0$, one retrieves consistently
\begin{align*}
&M_*^0\rightarrow m_B^0\ ,\quad A_*^0\rightarrow\alpha_B^0\ ,\\
&m_*^0\rightarrow m_A^0\ ,\quad\alpha_*^0\rightarrow\alpha_A^0\ ,\quad \beta_*^0\rightarrow\beta_A^0\ ,\quad {\beta'}_*^0\rightarrow{\beta'}_B^0\ ,
\end{align*}
 that is, $A$ becomes a test body orbiting around the central body $B$.\\
 
We note also that $\nu$-deformations do not enter the coefficients $b$ and $v_1$ in the generic $\nu\neq 0$ case, see (\ref{coeffsEffectifsJust1}), which are hence particularly simple~; we hence recover a feature of the \textit{linearized} effective dynamics which is common with that of the general relativity case (see Buonanno and Damour in \cite{Buonanno:1998gg}), and which is related to the very specific form of the quadratic functional relation (\ref{EOBquadrRel}).\footnote{We also recall that the gravitational coupling $G_{AB}=1+\alpha_A^0\alpha_B^0$, appearing in the two-body Hamiltonian (see (\ref{ham2corpsStructure}), subsection \ref{subsectionTwo_body_Ham}, and paper 1 section III C), encompasses the linear addition of the metric and scalar interations at linear level \cite{Damour:1995kt}. The present mapping has consistently split it again, between the effective metric and scalar sectors, i.e. $b$ and $v_1$, see (\ref{coeffsEffectifsJust1}), contrarily to the GR-centered, fully \textit{metric} mapping of paper 1, where $G_{AB}$ appeared at each post-Keplerian orders in the form $(G_{AB}M)/r$, $r$ being the Schwarzschild-Droste coordinate used there.\\}

\subsubsection{General relativity}
Finally, in the general relativity limit (\ref{GR_limit}), (\ref{effectmu}) and (\ref{effectM}) become the well-known reduced and total masses $m_*(\varphi)=\mu$ and $M_*(\varphi)=M$, and the effective coefficients (\ref{coeffsEffectifsJust}) reduce to
\begin{subequations}
\begin{align}
&a=2 \sqrt{1+6\nu}\ ,\quad b=2\ ,\\
&v_1=v_2=v_3=0\ .
\end{align}
\label{coeffEffectJustRG}%
\end{subequations}
In other words, $V_e=1$, i.e. the effective scalar field effects disappear.
The (non-perturbative) metric sector is now written in Just coordinates 
 and differs from the results of Buonanno and Damour \cite{Buonanno:1998gg} who worked out their analysis in Schwarzschild-Droste coordinates. In the present paper, we hence have on hands a resummation of the 2PN general relativity dynamics that differs from the one explored in \cite{Buonanno:1998gg}. The comparison and consistency of the two shall be commented upon subsection \ref{subsec_JFBD}.
  When moreover $\nu=0$, $a=b$ and the metric consistently reduces to Schwarzschild's, see comment below (\ref{constraintABQ}).

\subsection{ST-EOB dynamics}
Inverting the EOB mapping relation (\ref{EOBquadrRel}) yields the ``EOB Hamiltonian",
\begin{equation}
H_{\rm EOB}=M\sqrt{1+2\nu\left(\frac{H_e}{\mu}-1\right)}\ ,\quad\text{where}\quad\frac{H_e}{\mu}=\sqrt{D_eV_e+D_e^2\hat p_\rho^2+\frac{D_e}{C_e}\left(\frac{\hat p_\phi}{\hat\rho}\right)^2}\ ,
\label{inverseRelationHam}
\end{equation}
[where $D_e$, $C_e$ and $V_e$ are given in (\ref{rappel_D_C_V}) and (\ref{coeffsEffectifsJust})] which defines a resummation of the two-body 2PK Hamiltonian, $H$.\footnote{We recall that by construction, when restricted to 2PK level, $H_{\rm EOB}$ yields a dynamics which is canonically equivalent to that derived from $H$.\\} In the following we focus on some features of the resultant resummed dynamics, in the strong field regime. Hence and from now on, the 2PK-truncated function $V_e$ is to be considered as exact, along with $D_e$ and $C_e$.
\subsubsection{Effective dynamics}
As we shall see, the ST-EOB dynamics will follow straightforwardly from that derived from the effective Hamiltonian $H_e$. This can be obtained from Hamilton's equations ($\dot q=\partial H_e/\partial p$, $\dot p=-\partial H_e/\partial q$), or, as already remarked below (\ref{effectiveProblemJust}), can be equivalently interpreted as a geodesic of the (body-dependent) conformal metric $\tilde g_{\mu\nu}=V_eg^e_{\mu\nu}$~:
\begin{align}
&\hspace*{1cm}\,d\tilde s_e^{\,2}\equiv -D_eV_e\,dt^2+\frac{V_e}{D_e}d\rho^2+C_eV_e\,\rho^2d\phi^2\ .
\label{metriqueConfEffect}
\end{align}

The staticity and spherical symmetry of this metric imply the conservation of the energy and angular momentum of the orbit (per unit mass $\mu$),
\begin{align}
\quad u_t=-D_eV_e\frac{dt}{d\lambda}\equiv-E\ ,\quad u_\phi=C_eV_e\rho^2\frac{d\phi}{d\lambda}\equiv L\ ,\label{conservEandL}
\end{align}
$\lambda$ being an affine parameter along the trajectory.
When moreover the 4-velocity is normalized as $u^\mu u_\mu=-\epsilon$
(where $\epsilon=1$ for $\mu\neq 0$, $\epsilon=0$ for null geodesics), the radial motion is driven by an effective potential $F_\epsilon$\,,
\begin{equation}
\left(\frac{d\rho}{d\lambda}\right)^2=\frac{1}{V_e^2}F_\epsilon(u)\ ,
\label{EOMeffectJust}
\end{equation}
where
\begin{align}
&F_\epsilon(u)\equiv E^2-D_eV_e\left(\epsilon+\frac{j^2u^2}{C_eV_e}\right)\ ,\quad j\equiv\frac{L}{M}\ ,\quad u\equiv\frac{1}{\hat \rho}=\frac{M}{\rho}\ ,\\
\text{and}\qquad D_e(u)=&(1-au)^{b/a}\ ,\quad C_e(u)=(1-au)^{1-b/a}\ ,\quad V_e(u)=1+v_1u+v_2u^2+v_3u^3\ .\nonumber
\end{align}
\subsubsection{ISCO location}
We now focus on circular orbits when $\epsilon=1$, i.e., $F_{\epsilon=1}(u)=F'_{\epsilon=1}(u)=0$~; $j^2$ and $E$ are therefore related to $u$ through
\begin{equation}
j^2(u)=-\frac{(D_eV_e)'}{(u^2D_e/C_e)'}\ ,\quad
E(u)=\sqrt{D_eV_e\left(1+\frac{j^2(u)u^2}{C_eV_e}\right)}\ .\label{Ej}
\end{equation}
A characteristic feature of the strong-field regime is the innermost stable circular orbit (ISCO), which is reached when the third (inflection point) condition is satisfied $F''_{\epsilon=1}(u)=0$, i.e. when $u_{\rm ISCO}$ is the root, if any, of the equation~:
\begin{equation}
F'_{\epsilon=1}(u_{\rm ISCO})=F''_{\epsilon=1}(u_{\rm ISCO})=0\quad\Rightarrow\quad\frac{(D_eV_e)''}{(D_eV_e)'}=\frac{(u^2D_e/C_e)''}{(u^2D_e/C_e)'}\ .\label{ISCO}
\end{equation}
\subsubsection{Light-ring location}
When $\epsilon=0$, $F_{\epsilon=0}(u)=E^2-j^2u^2\frac{D_e}{C_e}$ and one can define a light-ring (LR), i.e. the radius of null circular orbits, through $F'_{\epsilon=0}(u_{\rm LR})=0$~:
\begin{equation}
u_{LR}=\frac{1}{b+\frac{a}{2}}\quad \Leftrightarrow\quad \rho_{LR}=M\left(2+\mathcal{R}\right)\ ,
\end{equation}
where $\mathcal R$ is given in (\ref{coeffsEffectifsJust}).
In particular one retrieves $\mathcal R=1$, i.e. $\rho_{LR}=3M$ (Schwarzschild's LR location) in the test-mass ($\nu\rightarrow0$), general relativity limit (\ref{GR_limit}).
\subsubsection{ST-EOB orbital frequency}
We now turn to the resummed two-body dynamics defined by the EOB Hamiltonian (\ref{inverseRelationHam}). Since $H_{\rm EOB}$ and $H_e$ are conservative, we have~:
\begin{equation}
\left(\frac{\partial H_{\rm EOB}}{\partial H_e}\right)=\frac{1}{\sqrt{1+2\nu(E-1)}}\label{timeRescaling}
\end{equation}
since $H_e=\mu E$ is a constant on-shell.
Therefore, the resummed equations of motion
\begin{equation}
\frac{d\rho}{dt}=\frac{\partial H_{\rm EOB}}{\partial p_\rho}\ ,\quad\frac{d\phi}{dt}=\frac{\partial H_{\rm EOB}}{\partial p_\phi}\ ,\quad\frac{dp_\rho}{dt}=-\frac{\partial H_{\rm EOB}}{\partial\rho}\ ,\quad\frac{dp_\phi}{dt}=-\frac{\partial H_{\rm EOB}}{\partial\phi}=0\ ,
\end{equation}
are identical to the effective ones, i.e. derived from the effective Hamiltonian, $H_e(q,p)$, to within the (constant) time rescaling $t\rightarrow t\sqrt{1+2\nu(E-1)}$.
In particular, for circular orbits, the orbital frequency reads
\begin{equation}
\Omega(u)\equiv\frac{d\phi}{dt}=\frac{\partial H_{\rm EOB}}{\partial H_e}\frac{\partial H_e}{\partial p_\phi}=\frac{D_e}{C_e}\frac{ju^2}{ME\sqrt{1+2\nu(E-1)}}\ ,
\label{Omega}
\end{equation}
where $E(u)$ and $j(u)$ are given for circular orbits in (\ref{Ej}). Its ISCO value is reached when $u=u_{\rm ISCO}$, as defined in (\ref{ISCO}).\\

Note that the orbital frequency has been derived in the Just coordinate system, $(q,p)$, which is related to the real one, $(Q,P)$, through the canonical transformation presented in subsection \ref{section_transfoCano}. Moreover, for circular orbits ($p_\rho=P_R=0$), $\Phi=\phi$,  and hence (\ref{Omega}) is the \textit{observed} orbital frequency. See also subsections \ref{section_transfoCano} and \ref{subsec_JFBD}.

\subsection{An example : the Jordan-Fierz-Brans-Dicke theory\label{subsec_JFBD}}
\subsubsection{A simple one-parameter model}
We now illustrate the previous results through the example of the Jordan-Fierz-Brans-Dicke theory \cite{Fierz:1956zz}, \cite{Brans:1961sx}, which depends on a unique parameter $\alpha$, such that\footnote{For a comparison with the Jordan-frame parameter $\omega$, such that $3+2\omega=\alpha^{-2}$, see \cite{Julie:2017pkb} appendix A.}
\begin{align}
&S_{\rm JFBD}[ g_{\mu\nu}, \varphi,\Psi]=\frac{1}{16\pi}\int d^4x\sqrt{-g}\,\bigg(R-2g^{\mu\nu}\partial_\mu\varphi\partial_\nu\varphi\bigg)+S_{m}\left[\Psi,\mathcal{A}^2 (\varphi) g_{\mu\nu} \right]\ ,\nonumber\\
&\hspace*{3cm}\text{where}\quad \mathcal A(\varphi)=e^{\alpha\varphi}\ ,\quad \alpha=\frac{d\ln \mathcal A}{d\varphi}=cst\ ,
\end{align}
while general relativity is retrieved when $\alpha=0$.\\

The two-body dynamics is then described by replacing $S_m$ by its ``skeleton" version,
\begin{equation}
S_m^{skel}[x_A^\mu,g_{\mu\nu},\varphi]=-\sum_A\int m_A(\varphi)\,ds_A\ ,
\end{equation}
where, for the sake of simplicity, we shall further neglect self-gravity effects, i.e.
$m_A(\varphi)=\mathcal A(\varphi)\,\tilde m_A$, where $\tilde m_A$ are constants, see discussion above (\ref{defSensitivities}). In that case, since $\mathcal A(\varphi)$ is known and the Jordan masses $\tilde m_A$ are constants, there is no need to expand $m_A(\varphi)$ as in (\ref{massExpansion}) since it is entirely determined as
\begin{equation}
m_A(\varphi)=m_A^0e^{\alpha(\varphi-\varphi_0)}\ , \qquad m_A^0=cst\ .
\end{equation}

Therefore, the fundamental parameters (\ref{defSensitivities}) become universal (\ref{limit_noSelfGrav}) and reduce to
\begin{align}
\alpha_A=\frac{d\ln m_A}{d\varphi}=\alpha\ ,\quad\beta_A=0\ ,\quad\beta'_A=0\ ,
\end{align}
and the post-Keplerian (two-body) parameters (\ref{defParamPK}) greatly simplify as well to
\begin{align}
G_{AB}=1+&\alpha^2\ ,\quad\bar\gamma_{AB}=-\frac{2\alpha^2}{1+\alpha^2}\ ,\quad\delta_A=\delta_B=\frac{\alpha^2}{(1+\alpha^2)^2}\ ,\nonumber\\
&\bar\beta_A=\bar\beta_B=0\ ,\quad\epsilon_A=\epsilon_B=0\ ,\quad\zeta=0\ .\label{JFBDsensitivities}
\end{align}
Hence, the coefficients (\ref{coeffsEffectifsJust}) of the functions
\begin{align}
D_e=\left(1-\frac{a}{\hat\rho}\right)^{\frac{b}{a}},\quad C_e=\left(1-\frac{a}{\hat\rho}\right)^{1-\frac{b}{a}},\quad
V_e=1+\frac{v_1}{\hat{\rho}}+\frac{v_2}{\hat{\rho}^2}+\frac{v_3}{\hat{\rho}^3}+\cdots\ ,\label{rappel_V}
\end{align}
depend only on $\alpha$ and $\nu=\mu/M$ and reduce to
\begin{subequations}
\begin{align}
&b=2\ ,\qquad v_1=-2\alpha^2\ ,\\
&a=2\mathcal{R}\ ,\quad 
v_2=2\alpha^4-2\alpha^2\mathcal{R}\ ,\\
&v_3=\frac{4}{3}\alpha^2\bigg(3\alpha^2\mathcal{R}-(2+2\alpha^2+\alpha^4)-\nu(14+12\alpha^2-2\alpha^4)\bigg)\ ,\\
\text{with}\quad &\mathcal{R}(\nu)=\sqrt{(1+\alpha^2)\bigg(1+2(3-\alpha^2)\nu\bigg)}\ .\nonumber
\end{align}\label{coeffsEffectJustBrans}
\end{subequations}

\subsubsection{An improved $V_e$ function}
As discussed in subsection \ref{subsectionMapping}, the effective dynamics is a $\nu$-deformation of a ST test-body problem, which, in the present case, describes a test particle $m_*(\varphi)=\mu\, e^{\alpha(\varphi-\varphi_0)}$ orbiting around a central body $M_*(\varphi)=M\,e^{\alpha(\varphi-\varphi_0)}$, where $\mu=m_A^0m_B^0/M$ and $M=m_A^0+m_B^0$, see (\ref{nouveauxParamEffectifs}) and below.\\

 Therefore, in keeping with our approach consisting in centering as much as possible the effective dynamics on the test-body problem, we can ``improve" $V_e$ by factorizing out its exact, $\nu=0$ expression~:
\begin{equation}
V_e= V_{exact}^{\nu=0}P(\nu)\ ,\quad P(\nu)=1+\frac{p_1}{\hat\rho}+\frac{p_2}{\hat\rho^2}+\frac{p_3}{\hat\rho^3}+\cdots\ ,\label{V_factorized}
\end{equation}
where, by definition, see (\ref{def_V}),
\begin{equation}
V_{exact}^{\nu=0}\equiv \left(\frac{m_*(\varphi_e)}{m_*^0}\right)^2=e^{2\alpha\varphi_e}\ ,
\end{equation}
and where $\varphi_e$ is the scalar field generated by the central body, see (\ref{param_a_b_oneBody})~:
\begin{equation}
\varphi_e=\varphi_0+\frac{\alpha}{2\sqrt{1+\alpha^2}}\ln\left(1-\frac{2\sqrt{1+\alpha^2}}{\hat\rho}\right)\ ,\quad \hat{\rho}=\rho/M\ .\label{phiE_jfbd}
\end{equation}

 The 2PK identification of (\ref{V_factorized})-(\ref{phiE_jfbd}) with (\ref{rappel_V})-(\ref{coeffsEffectJustBrans}) gives then
\begin{align}
&V_e=\left(1-\frac{2\sqrt{1+\alpha^2}}{\hat\rho}\right)^{\frac{\alpha^2}{\sqrt{1+\alpha^2}}}P(\nu)\ ,\quad P(\nu)=1+\frac{p_1}{\hat\rho}+\frac{p_2}{\hat\rho^2}+\frac{p_3}{\hat\rho^3}\ ,\label{VeJFBD}\\
\text{with}\quad& p_1=0\ ,\quad p_2=2\alpha^2\left[\mathcal{R}(0)-\mathcal{R}(\nu)\right]\ ,\quad p_3=-\frac{8}{3}\alpha^2(7+6\alpha^2-\alpha^4)\nu\ ,\nonumber
\end{align}
where $P(\nu=0)=1$. In doing so, in the test-mass limit, $D_e$, $C_e$ \textit{as well as} $V_e$ reduce to their \textit{exact}, non perturbative expressions, to which they are smoothly connected.

\subsubsection{The ST-EOB orbital frequency at the ISCO}
\begin{figure}[h!]
\centering
\caption{ISCO location (left panel) in Just coordinates and ISCO frequency (right panel) versus the (squared) Jordan-Fierz-Brans-Dicke parameter $\alpha^2$, when $\nu=0$ (dashed lines) and $\nu=0.25$ (solid lines).
}
\begin{subfigure}{.5\textwidth}
  \centering
  \includegraphics[width=0.95\linewidth]{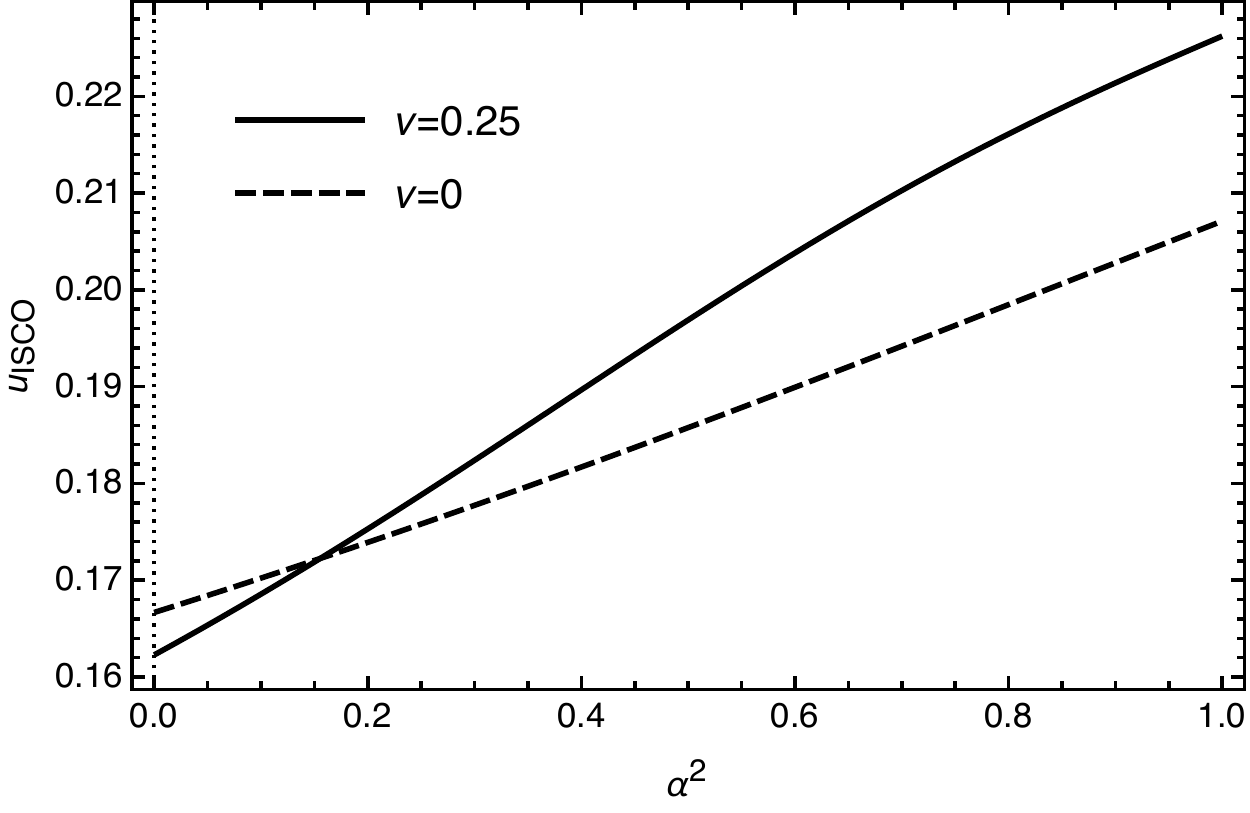}
  \label{fig:sub1}
\end{subfigure}%
\begin{subfigure}{.5\textwidth}
  \centering
  \includegraphics[width=0.95\linewidth]{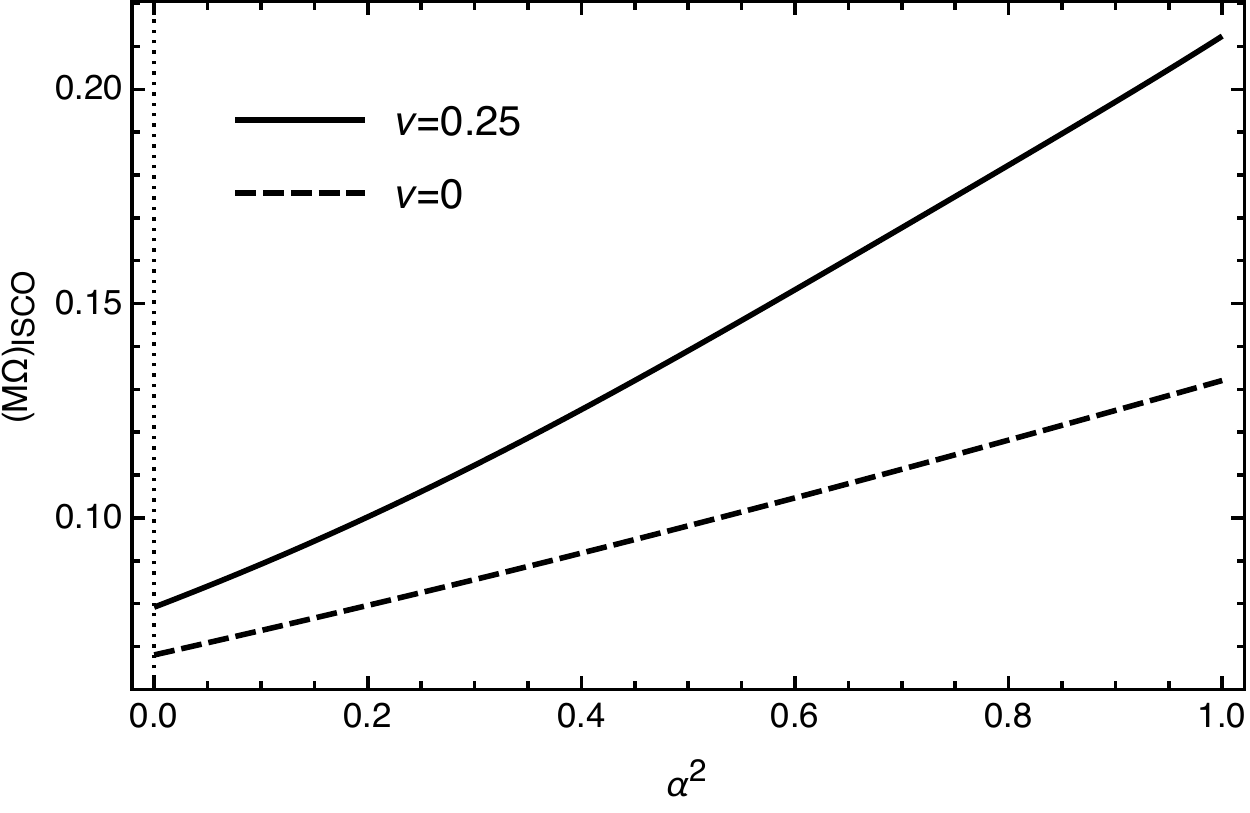}
  \label{fig:sub2}
\end{subfigure}
\label{figureJFBD}
\end{figure}

We now have on hands all the necessary material to study the ISCO location, $u_{ISCO}\equiv M/\rho_{ISCO}$, and associated orbital frequency, $M\Omega_{ISCO}$, as defined in the previous subsection, using (\ref{Ej}), (\ref{ISCO}) and (\ref{Omega}). The results are even in $\alpha$, as expected from (\ref{coeffsEffectJustBrans}) and (\ref{VeJFBD}) and are gathered in figure \ref{figureJFBD}, for $0<\alpha^2<1$.\\

The limit $\alpha=0$ reduces to general relativity. When moreover $\nu=0$, one recovers the well-known Schwarzschild values $u_{ISCO}=1/6$, $M\Omega_{ISCO}=0.06804$ (since then the Just and Droste-Schwarzschild coordinates coincide, see comment below \ref{coeffEffectJustRG}). Note that when $\alpha=0$ but $\nu\neq 0$, $u_{ISCO}$ is \textit{less} than $1/6$. This does not contradict the general relativity results of Buonanno and Damour \cite{Buonanno:1998gg}, who worked in Droste coordinates rather than Just's~; 
 rather, this illustrates the fact that the effective radii are physically irrelevant, contrarily to the orbital frequency $M\Omega_{ISCO}$ which \textit{is} an observable~:
for $\alpha=0$ and for all $\nu\neq 0$, the ISCO frequency turns out to be always larger than the Schwarzschild one (see right panel of figure \ref{figureJFBD}), as in \cite{Buonanno:1998gg}. For instance, when $\nu=1/4$, we find $M\Omega_{ISCO}=0.07919$, i.e. slightly higher than the value $0.07340$ quoted in \cite{Buonanno:1998gg}. The $\sim 7\%$ difference in the numerical values is reasonable considering that the two resummations (see (\ref{coeffEffectJustRG})) are different and built on 2PK information only.

\vfill\eject

Now, when $\alpha\neq0$, i.e. when the scalar field is switched on, the ISCO frequency increases roughly linearly in $\alpha^2$, as can be seen from the right panel of figure \ref{figureJFBD}, with a slope
\begin{equation}
\left.\frac{d(M\Omega_{ISCO})}{d(\alpha^2)}\right|_{\nu=1/4}\simeq 0.13\quad\text{and}\quad\left.\frac{d(M\Omega_{ISCO})}{d(\alpha^2)}\right|_{\nu=0}\simeq 0.063\ .\label{sensitivityJFBD}
\end{equation}

Interestingly, when restricted to a perturbative regime $\alpha<<1$, these results are qualitatively consistent with the ones obtained from the \textit{distinct}, GR-centered resummation of \cite{Julie:2017pkb}, where ST effects were considered as perturbations of general relativity. There, we started from the best available EOB-NR metric, known in GR at 5PN order, see \cite{Damour:2014sva}, \cite{Nagar:2015xqa}, and \cite{Bini:2013zaa}. We then perturbed this effective metric by scalar-tensor 2PK corrections and studied their impact on the strong field dynamics. The ISCO frequency was also found there to increase linearly with the ``PPN", Eddington parameter
\begin{equation}
\epsilon_{1PK}\equiv \langle\bar\beta\rangle-\bar\gamma_{AB}
\end{equation}
 (which reduces to $\epsilon_{1PK}\sim 2\alpha^2$ in the present case, see (\ref{defParamPK2}), (\ref{meanOverBodiesParam}) and (\ref{JFBDsensitivities})), the slope being numerically of the same order of magnitude,  hence illustrating the robustness of the EOB description of the strong field regime.\footnote{In particular, we found $d(G_{AB}M\Omega)/d\epsilon_{1PK}\simeq 0.13$ in the equal-mass case. In the present paper we will not proceed to any detailed, quantitative comparison of the two resummations since the present ST-centered approach is limited in this section to the JFBD case and since paper 1 included some extra 5PN GR information.\\
 }\\

More importantly, we have developped, throughout this paper, a ST-centered EOB Hamiltonian that reduces to the \textit{exact} test-body Hamiltonian in the test-mass limit.
In consequence, the ISCO predictions are well-defined even when $|\alpha|\sim 1$, that is, can be pushed to a regime that strongly departs from general relativity~: there, the estimated ISCO location and frequency significantly deviate from the GR ones \textit{and} remain smoothly connected to the test-mass ($\nu=0$) limit (see figure \ref{figureJFBD}), which we know exactly even in the strong field regime.\footnote{It must be noted that when $\alpha>\alpha_{\rm crit}\simeq 1.6$, the exact test-body problem (which is reached when $\nu=0$) does not feature any ISCO anymore, since then (\ref{ISCO}) has no root. This phenomenon is encompassed by our mapping~; when $\nu$ is non zero and increases, the value of $\alpha_{\rm crit}$ smoothly decreases to reach $\alpha_{\rm crit}(\nu=1/4)\simeq 1.03$.\\}\\

We hence have illustrated, in the simple case of the Jordan-Fierz-Brans-Dicke theory, the \textit{complementarity} of two EOB resummations of the scalar-tensor dynamics~:\\ \\
(i) The first one, introduced in paper 1, which is built on rich (5PN) general relativity information, is oriented towards regimes where ST effects are considered as perturbations of GR [while the dynamics is ill-defined in non-perturbative regimes~; this necessitates, e.g., the use of appropriate Pad\'e resummations of the ST perturbations as soon as $\epsilon_{1PK}\gtrsim 10^{-1}$, see \cite{Julie:2017pkb} for details].\\ \\
(ii) The second, ST-centered one, that we have developped throughout this paper, which
has been shown to be well-suited to describe regimes that may depart strongly from general relativity~; the price to pay being that it is based on 2PK information only.

\section{Concluding remarks}

The reduction to a simple, effective-one-body motion has been a key element in the  treatment of the two-body problem in general relativity. In the pionnering 1998 paper \cite{Buonanno:1998gg} of Buonanno and Damour, the 2PN effective dynamics was found to be a $\nu$-deformation of the test-body problem in GR, namely, the geodesic motion of a test particle $\mu$ in the Schwarzschild metric generated by a central body $M$.\\

Remarkably, the fruitfulness of the EOB approach spreads beyond the scope of general relativity~: indeed, by means of a canonical transformation and the same EOB quadratic relation (\ref{EOBquadrRel}), we reduced the 2PK two-body dynamics in scalar-tensor theories to a $\nu$-deformed version of the ST test-body problem~; namely, the motion of a 
test particle [$\mu$, $\alpha_*^0$, $\beta_*^0$, ${\beta'}_*^0$] orbiting in the fields of a central body [$M$, $A_*^0$].\\

The present mapping has led, just like that of paper 1 \cite{Julie:2017pkb}, to a much simpler and compact description of the two-body dynamics in the 2PK regime, ``gauging away" the irrelevant information in a canonical transformation.  The (conservative) dynamics derived from the two ST-EOB Hamiltonians presented in \cite{Julie:2017pkb} and in the present paper are, by construction, canonically equivalent at 2PK order but, when taken as being exact, they define two distinct resummations of the dynamics in the strong field regime. The fact that both lead to consistent ISCO predictions (in their overlapping ST regimes) is a hint that they may have captured accurately some of the strong field features of binary coalescence in ST theories.\\

To summarize, we have on hands two complementary EOB dynamics :
(i) the geodesic motion in an effective metric in Schwarschild-Droste coordinates, encompassing the most accurate (5PN) GR information, which is particularly well-suited to test scalar-tensor theories when considered as parametrised corrections to general relativity \cite{Julie:2017pkb}, and (ii) a ST effective test-body problem, in Just coordinates, that allows to investigate regimes that depart strongly from GR, as was illustrated by the JFBD example (see subsection \ref{subsec_JFBD}). An exhaustive study of generic ST theories (that depend on five parameters (\ref{effectiveProblemJust})) is left to future works.
Note that one cannot perform the 2PK Droste-Just coordinate change $r^2=C_eV_e\rho^2$ without spoiling either the resummation towards the ST test-body problem of (ii) or the 5PN accurate GR information of (i).\\

Now, Solar System and binary pulsar experiments have already put stringent constraints on ST theories, namely,  $(\alpha_A^0)^2<4\times 10^{-6}$ for any body $A$, and $\alpha^2<2\times 10^{-5}$ in (non self-gravitating) JFBD theory, see, e.g., \cite{Freire:2012mg} and \cite{Wex:2014nva}). Since the parameters (\ref{paramEffectFund}) contain terms that are all driven by at least $(\alpha_{A/B}^0)^i$, $i\geq 2$, these constraints seem to imply that scalar-tensor effects are negligible.
However, gravitational wave astronomy allows to observe
new regimes of gravity that might escape these constraints. For example, stars that are subject to dynamical scalarization \cite{Barausse:2012da} can develop nonperturbative $\alpha_A^0$ parameters during the few last orbits before plunge (they can numerically reach order unity \cite{Palenzuela:2013hsa}), that is, in the strong field regime which is \textit{precisely} explored by our EOB approach. Also, from the cosmological point of view, GR is indeed an attractor of ST theories \cite{Damour:1992kf}, \cite{Damour:1993id}, and hence,
gravitational wave detectors, which are designed to observe sources at high redshifts can probe epochs when ST effects may have been stronger.
Hence, the tools developped in the present paper, which goes beyond the scope of \cite{Julie:2017pkb}, could turn out to become useful in practice.\\

Finally, we recall that SSS black holes cannot carry scalar hair in the class of ST theories we are considering here (provided that the no hair theorems hold in the highly dynamical regime of a merger), see, e.g., the comments below equation (\ref{constraintABQ}) and references quoted there.
An interesting alternative would be to induce hair by means of a massless gauge vector field, as for, e.g., Einstein-Maxwell-Dilaton theories \cite{Jai-akson:2017ldo} \cite{Hirschmann:2017psw}, which will be the subject of future works.

\section*{Acknowledgements}
I am very grateful to Nathalie Deruelle who guided me throughout this project, and who carefully read and commented the manuscript of the present paper.

\vfill\eject

\appendix

\section{Canonically-transformed effective Hamiltonians\label{appendix_canoTransf_He}}
Performing the canonical transformation (\ref{transfoCano}-\ref{generatrice}), the effective 2PK Hamiltonian (\ref{HamEffectJust2PK}) is rewritten in the intermediate coordinate system $(q,p)\rightarrow(Q,p)$~:
\begin{equation*}
\hat H_e=1+\left(\frac{\mathcal P^2}{2}+\frac{v_1-b}{2\hat R}\right)+\hat H_e^{\rm 1PK}+\hat H_e^{2\rm PK}+\cdots
\end{equation*}
where
\begin{align*}
&\hat H_e^{1PK}=\hat p_r^4 \bigg(2 \alpha _1+3 \beta _1\bigg)-\hat p_r^2 {\cal P}^2 \bigg(\alpha_1+3 \beta _1\bigg)+{\cal P}^4 \left(-\alpha _1-\frac{1}{8}\right)\\
&+\frac{1}{4\hat R}\bigg[ {\cal P}^2 \bigg(2 a+2 \alpha _1 b-3 b-4 \gamma _1-\left(2 \alpha _1+1\right) v_1\bigg)-2 \hat p_r^2 \bigg(a-2 \alpha _1 \left(b-v_1\right)-3 \beta _1 \left(b-v_1\right)-2 \gamma _1\bigg)\bigg]\\
&+\frac{1}{8\hat R^2}\bigg[b \left(-2 a+b+4 \gamma _1\right)-2 v_1 \left(b+2 \gamma _1\right)-v_1^2+4 v_2\bigg]\quad ,
\end{align*}

\begin{align*}
&\hat H_e^{2PK}=-\frac{1}{2} \hat p_r^6 \bigg(36 \alpha _1 \beta _1+12 \alpha _1^2+27 \beta _1^2-4 \beta _2-10 \gamma _2\bigg)+\frac{1}{2} \hat p_r^4 {\cal P}^2 \bigg(2 \alpha _1 \left(9 \beta _1-1\right)+8 \alpha _2+27 \beta _1^2-3 \beta _1+2 \beta _2-10 \gamma _2\bigg)\\
&+\frac{1}{2} \hat p_r^2 {\cal P}^4 \bigg(\alpha _1 \left(18 \beta _1+1\right)+9 \alpha _1^2-6 \alpha _2+3 \beta _1-6 \beta _2\bigg)+\frac{1}{16} \bigg(24 \alpha _1^2+8 \alpha _1-16 \alpha _2+1\bigg) {\cal P}^6\\
&+\frac{1}{16\hat R}\bigg[8 \hat p_r^4 \bigg(2 \alpha _1 \left(3 \left(a-2 b \beta _1-b-2 \gamma _1\right)+\left(6 \beta _1-1\right) v_1\right)+3 \beta _1 \left(3 a-3 b-6 \gamma _1-v_1\right)+2 b \beta _2+5 b \gamma _2-4 \alpha _1^2 \left(b-v_1\right)\\
&-9 \beta _1^2 \left(b-v_1\right)+4 \delta _2-2 \beta _2 v_1-5 \gamma _2 v_1+6 \epsilon _2\bigg)\\
&-4 \hat p_r^2 {\cal P}^2 \bigg(-2 \alpha _1 \left(-3 a-6 b \beta _1+6 b+6 \gamma _1+\left(6 \beta _1+2\right) v_1\right)+18 a \beta _1-a-27 b \beta _1-6 b \beta _2-36 \beta _1 \gamma _1\\
&+8 \alpha _1^2 \left(b-v_1\right)-8 \alpha _2 \left(b-v_1\right)+2 \gamma _1+4 \delta _2-9 \beta _1 v_1+6 \beta _2 v_1+12 \epsilon _2\bigg)\\
&+{\cal P}^4 \bigg(\alpha _1 \left(-24 a+36 b+48 \gamma _1\right)-4 a-8 \alpha _1^2 b+8 \alpha _2 b+5 b+8 \gamma _1-16 \delta _2+\left(8 \alpha _1^2+12 \alpha _1-8 \alpha _2+3\right) v_1\bigg)\bigg]\\
&+\frac{1}{16\hat R^2}\bigg[{\cal P}^2 \bigg(8 a^2+8 \alpha _1 a b+v_1 \left(-4 a+8 \alpha _1 \left(b+2 \gamma _1\right)+6 b+12 \gamma _1-8 \delta _2\right)-18 a b-24 a \gamma _1-4 \alpha _1 b^2+9 b^2\\
&-16 \alpha _1 b \gamma _1+36 b \gamma _1+8 b \delta _2+24 \gamma _1^2-16 \eta _2-4 \left(4 \alpha _1+1\right) v_2+\left(4 \alpha _1+3\right) v_1^2\bigg)\\
&\left.-4 \hat p_r^2 \bigg(2 a^2-6 a b \beta _1-v_1 \left(a+4 \alpha _1 \left(b+2 \gamma _1\right)+6 \beta _1 \left(b+2 \gamma _1\right)-2 \gamma _1-4 \delta _2-6 \epsilon _2\right)+\alpha _1 \left(2 b \left(-2 a+b+4 \gamma _1\right)+8 v_2\right)\right.\\
&-3 a b-6 a \gamma _1+3 b^2 \beta _1+12 b \beta _1 \gamma _1+6 b \gamma _1-4 b \delta _2-6 b \epsilon _2+6 \gamma _1^2-4 \eta _2-v_1^2 \left(2 \alpha _1+3 \beta _1\right)+12 \beta _1 v_2\bigg)\bigg]\\
&+\frac{1}{48\hat R^3}\bigg[-3 v_1 \left(2 a b-b^2-8 \gamma _1 \left(b+\gamma _1\right)+8 \eta _2+4 v_2\right)-12 \left(b \gamma _1 \left(-2 a+b+2 \gamma _1\right)-2 b \eta _2+v_2 \left(b+4 \gamma _1\right)\right)\\
&-b (b-4 a) (b-2 a)+3 v_1^2 \left(b+4 \gamma _1\right)+3 v_1^3+24 v_3\bigg]\quad .
\end{align*}

\bibliographystyle{unsrt}

\end{document}